\documentclass[floatfix,aps,pra,longbibliography,twocolumn,a4paper,nofootinbib,notitlepage]{revtex4-1}
\usepackage[svgnames]{xcolor}
\usepackage[colorlinks=true,allcolors=magenta,breaklinks=true]{hyperref}
\usepackage[english]{babel}
\usepackage{amsmath,amsfonts,graphicx}
\usepackage{tikz}
\usetikzlibrary{positioning,calc,arrows,decorations.pathmorphing,shapes}

\begin{document}
\title{The Free Energy  of a General Computation}
\author{\"Amin Baumeler}
\affiliation{Institute for Quantum Optics and Quantum Information (IQOQI), Austrian Academy of Sciences, Boltzmanngasse 3, 1090 Vienna, Austria}
\affiliation{Facolt\`{a} indipendente di Gandria, Lunga scala, 6978 Gandria, Switzerland}
\author{Stefan Wolf}
\affiliation{Faculty of Informatics, Universit\`a della Svizzera italiana, Via Buffi 13, 6900 Lugano, Switzerland}
\affiliation{Facolt\`{a} indipendente di Gandria, Lunga scala, 6978 Gandria, Switzerland}

\begin{abstract}
\noindent
Starting from Landauer's slogan ``information is physical,'' we revise and modify Landauer's principle stating that the {\em erasure\/} of information has a minimal price in the form of a certain quantity of free energy. We establish a direct link between the {\em erasure cost\/} and the {\em work value\/} of a piece of information, and show that the former is essentially the length of the string's best compression by a reversible computation. 
We generalize the principle by deriving bounds on the free energy to be invested for~--- or gained from, for that matter~--- a~{\em general computation}.
We then revisit the {\em second law of thermodynamics\/} and compactly rephrase it (assuming the Church/Turing/Deutsch hypothesis that physical reality can be simulated by a universal Turing machine): Time evolutions are {\em logically reversible\/}~--- ``the future fully remembers the past (but not necessarily {\em vice versa\/}).''
We link this view to previous formulations of the second law, and we argue that it has a particular feature that suggests its ``logico-informational'' nature, namely {\em simulation resilience\/}: If a computation faithfully simulates a physical process violating the law~--- then that very computation procedure violates it as well.
\end{abstract}

\maketitle

\section{Introduction}
In 1961, {\em Rolf Landauer\/} famously stated ``Information is physical''~\cite{Landauer1998}: Despite the success of Shannon's making information an {\em abstract\/} concept (that can be viewed and understood independently of its particular  physical realization), Landauer~--- while  not questioning the power of that abstract view~--- recalls that all information storing, treatment, and transmission  is ultimately a  {\em physical\/} process and, thus, subject to physical laws.
A~specific law relevant in this context is the second law of thermodynamics.
Its consequence for information processing has been called {\em Landauer's principle\/}~\cite{Landauer1961}:
{\em ``The erasure of~$N$ bits of information costs at least an amount of $NkT\ln 2$ ($k$ being Boltzmann's constant) of free energy that must be dissipated as heat into the environment of temperature~$T$.''\/}
(Note that this heat dissipation  is crucial for the argument:
It represents the compensation required for avoiding a violation of the second law despite an entropy decrease in the memory device through the erasure process.)
Conversely, erased strings have a {\em work value\/} (see, {\it e.g.}, Ref.~\cite{Szilard1929,Dahlsten2009}):
By, {\it e.g.}, encoding an erased bit string of length~$N$ in the particles position of a gas within a box, where the particle's position is on the left half for the value~$0$ and on the right half otherwise, and by placing a piston in the center,~$NkT\ln 2$ of free energy can be extracted from the environment, ``randomizing'' the original string.

In this article, we modify and generalize Landauer's principle in the following respects:
First, it is claimed that the erasure cost is not proportional to the length of the string to be erased, but of its best {\em compression\/} --- given the entire knowledge of the erasure device (Section~\ref{sec:revlandauer}).
We obtain these results from new bounds on the {\em work value\/} of information (Section~\ref{sec:data}), and a direct connection between erasure cost and work value of any~piece of information. Second, we generalize these results to a~lower bound on the free-energy cost, or value, of a {\em general\/} computation (Section~\ref{sec:gen}).
Our findings are modifications of known results (see {\it e.g.,} Refs.~\cite{Bennett1982,Dahlsten2009,Dahlsten2011,Faist2015}) to the {\em constructive\/} setting --- where all involved processes are imagined to be carried out by a Turing machine. 
Furthermore, we give a lower bound on the free-energy {\em gain\/} possible from certain computations~--- a~bound that  matches the {\em cost\/} of the inverse computation.
We look at the use of the erasure cost as an {\em intrinsic randomness\/} measure in the context of quantum correlations  (Section~\ref{sec:quantum}).
Having these results at hand, we finish by proposing a computational version of the second law of thermodynamics (Section~\ref{sec:logrev}).
This comes with a speculation about what trait of it is the reason why such a version exists in the first place.
Candidates are its ``encoding independence'' and ``simulation resilience:''
If a computation simulates a process violating the second law, then that computing procedure  {\em cannot  be closed\/}   but must dissipate ``junk'' bits onto the other parts of the tape,  or heat into the environment.
Thus, the violation of the law by a~process  carries over to {\em its simulation}.
The reason is that a  degree of freedom is represented  by~---  a~degree of freedom.

\section{Work Value: State of the Art}
While other results (see, {\it e.g.}, Refs.~\cite{PV,Sagawa2012,Deffner2013,Merhav2017,Boyd2018,Frank}) on the work value of information focus on using information reservoirs to generate energy flows, the below described results and this article focus on the work value of information --- being in form of random variables or bit strings --- {\it per se}.
As opposed to discussing the role of information in thermodynamic processes, we discuss the thermodynamics processes of information.

\subsection{Bennett's view}
Bennett~\cite{Bennett1982} claimed the work value of a string~$S$,~${\rm WV}(S)$, to be proportional to the difference between its length,~${\rm len}(S)$, and the length of the shortest program that produces~$S$.
The latter is called the {\em Kolmogorov complexity of~$S$}, denoted by~$K(S)$~\cite{Kolmogorov1965}.
Expressed mathematically, this amounts to
\begin{align}
	{\rm WV}(S)=({\rm len}(S)-K(S))kT\ln 2
	\,.
\end{align}
Bennett's argument is that~$S$ can be logically, hence, thermodynamically~\cite{Fredkin1982} {\em reversibly\/} mapped to the string $P||000\cdots 0$, where the symbol~$||$ denotes concatenation, and~$P$ is the shortest program generating $S$.
The length of the generated $\mbox{$0$-string}$ is ${\rm len}(S)-K(S)$.

It was already pointed out by Zurek~\cite{Zurek1989} that while it is true that the reverse direction exists and is computable by a universal Turing machine, its {\em forward direction}, {\em i.e.},~obtaining~$P$ from~$S$, is {\em uncomputable}.
This means that a ``demon'' that could carry out this work-extraction computation on~$S$ {\em does not exist\/} (if the Church/Turing hypothesis is true); the Kolmogorov complexity is an {\em uncomputable\/} value.
We will see, however, that Bennett's value is an {\em upper bound\/} on the work value of $S$. 
Bennett also links the string's erasure cost to its probabilistic entropy~\cite{Bennett2003}.

\subsection{Dahlsten {\em et al.}'s view}
Dahlsten {\em et al.}~\cite{Dahlsten2009,Dahlsten2011} follow Szil\'{a}rd~\cite{Szilard1929} in putting the {\em knowledge\/} of the demon extracting the work to the center of their attention.
More precisely, they claim ${\rm WV}(S)={\rm len}(S)-D(S)$, where the ``defect'' $D(S)$ is bounded from above and below by a {\em smooth R\'enyi entropy\/} of the distribution of $S$ from the demon's viewpoint, modeling its ignorance.
Building on these results and in the same probabilistic spirit, the cost of erasure~\cite{Rio2011} as well as of general computations~\cite{Faist2015} have been linked to entropic expressions of (conditional) probability distributions.

\section{Work Extraction as Data Compression}
\label{sec:data}
In the following, we model work extraction to be an algorithm executed by a ``demon with knowledge.'' 

\subsection{The Model}
We assume the demon to be 
a {\em universal Turing machine\/~${\cal U}$\/} the memory tape of which is
sufficiently long for the inputs and tasks in question, but {\em
  finite}.
The tape initially contains~$S$, the string the work value of which is
to be determined, $X$, a~finite string modeling the  demon's {\em knowledge 
about~$S$}, and~$0$s for the  rest of the tape. After  the
 extraction computation, the tape contains, at the bit positions 
initially holding $S$, a (shorter) string $P$ plus 
$
0^{{\rm len}(S)-{\rm len}(P)}
$,
whereas 
the rest of the tape
is (again) the same  as before the work extraction. 
The  operations are
{\em logically\/} reversible and can, hence, be carried out
{\em thermodynamically\/} reversibly~\cite{Fredkin1982}. 
Logical reversibility is the 
ability of the same demon to carry out the backward computation step
by step,
{\em i.e.}, from $P||X$ to~$S||X$.
We denote by~${\rm WV}(S|X)$   the {\em maximal 
length of an all-$0$-string 
extractable logically reversibly from $S$, given the knowledge~$X$},
times $kT\ln 2$,
{\em i.e.},
\begin{align}
	{\rm WV}(S|X):=({\rm len}(S)-{\rm len}(P))kT\ln 2
\end{align}
if $P$'s length is minimal.

\subsection{Lower Bound}
We show that every specific data-compression algorithm leads to a lower bound on the extractable work:
Let~$\mathcal{Z}$ be a computable function
\begin{align}
	\mathcal{Z}\, :\, \{0,1\}^*\times \{0,1\}^* \longrightarrow \{0,1\}^*
\end{align}
such that 
\begin{align}
	(A,B)\mapsto (\mathcal{Z}(A,B),B)
\end{align}
is injective.\footnote{The set~$\{0,1\}^*$ is the set of all finite but arbitrarily long bit strings.}
We call $\mathcal{Z}$ a {\em data-compression algorithm with helper}.
Then we have 
\begin{align}
	{\rm WV}(S|X)\geq ({\rm len}(S)-{\rm len}(\mathcal{Z}(S,X)))kT\ln 2\, .
	\label{eq:wvlowerbound}
\end{align}

\begin{figure}
	\centering
	\includegraphics[scale=0.5]{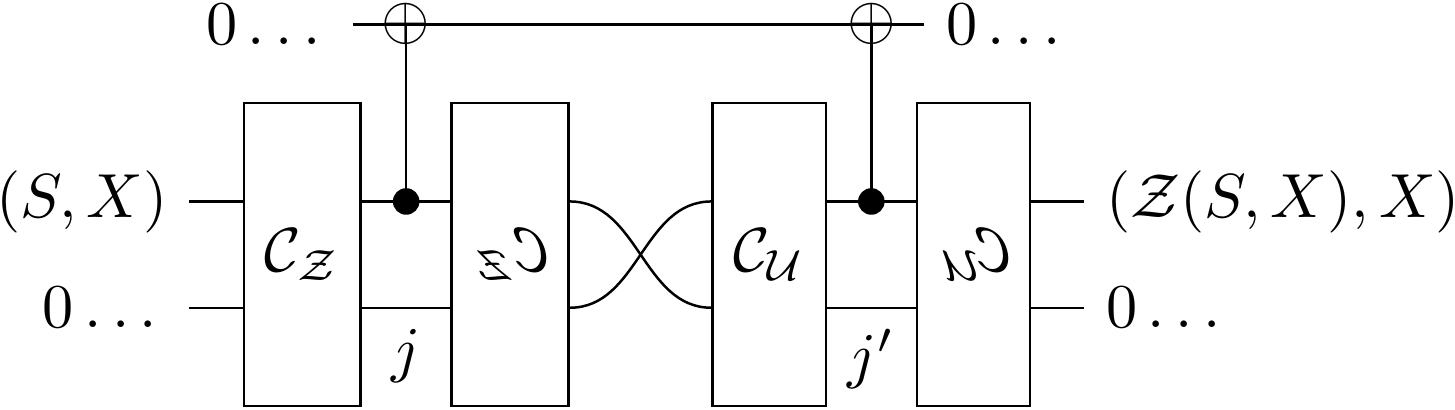}
	\caption{Schematic circuit of thermodynamically neutral compression with helper. The circuit~$\mathcal{C}_\mathcal Z$ implements the compression algorithm~$\mathcal{Z}$ with Toffoli gates only,~\protect\reflectbox{$\mathcal{C}_\mathcal Z$} is the same circuit in reverse order. Then again, the circuit~$\mathcal{C}_\mathcal U$ implements the corresponding {\em decompression\/} algorithm with Toffoli gates only, and~\protect\reflectbox{$\mathcal{C}_\mathcal U$} is its reverse. The symbols~$j$ and~$j'$ represent the ``junk'' that arises from implementing the circuits with Toffoli gates only.}
	\label{fig:bennett}
\end{figure}

This can be seen as follows.
First, note that the function
\begin{align}
	A||B\  \mapsto\ \mathcal{Z}(A,B)||0^{{\rm len}(A)-{\rm len}(\mathcal{Z}(A,B))}||B
\end{align}
is computable and bijective.
From the two (possibly irreversible) circuits which compute the compression and its inverse, one can obtain a {\em reversible\/} circuit for the function such that no further input or output bits are involved:
This can be achieved by first implementing all logical operations with Toffoli gates and uncomputing  the ``junk''~\cite{Bennett1973} in both circuits.
The resulting two circuits have now still the property that the input is part of the output. 
As a second step, we can simply combine the two such that the first circuit's first and second outputs become the second's second and first inputs, respectively.
Roughly speaking, the first computes the compression and the second reversibly uncomputes the raw data (see Figure~\ref{fig:bennett}).
The combined circuit has only {\em the compressed data plus the~$0$s\/} as the output, sitting on the bit positions carrying the input before. 
(This circuit is roughly as efficient as the less efficient of the two irreversible circuits for data compression and  decompression, respectively.) 
A typical example for an algorithm that can be used here is universal data compression {\em \`a la\/} Ziv-Lempel~\cite{Ziv1978}.

\subsection{Upper Bound}
We have the following upper bound on the extractable work:
\begin{align}
	{\rm WV}(S|X)\leq ({\rm len}(S)-K_{\cal U}(S|X))kT\ln 2\, ,
	\label{eq:wvupperbound}
\end{align}
where $K_{\cal U}(S|X)$ is the conditional Kolmogorov complexity (with
respect to the universal Turing machine ${\cal U}$\/) of $S$ given~$X$, {\em i.e.}, the length of the shortest program  $P$ for ${\cal
  U}$ 
that outputs~$S$, given~$X$. 
The reason is that  the extraction demon is only  able to carry out the
computation in question (logically, hence, thermodynamically)
reversibly 
 if it is able to carry
  out 
the reverse computation  as well.  Therefore, the string $P$ must
be at least as long as the shortest program for ${\cal U}$ generating~$S$ if~$X$ is given.

\begin{figure}
	\centering
	\includegraphics[scale=0.5]{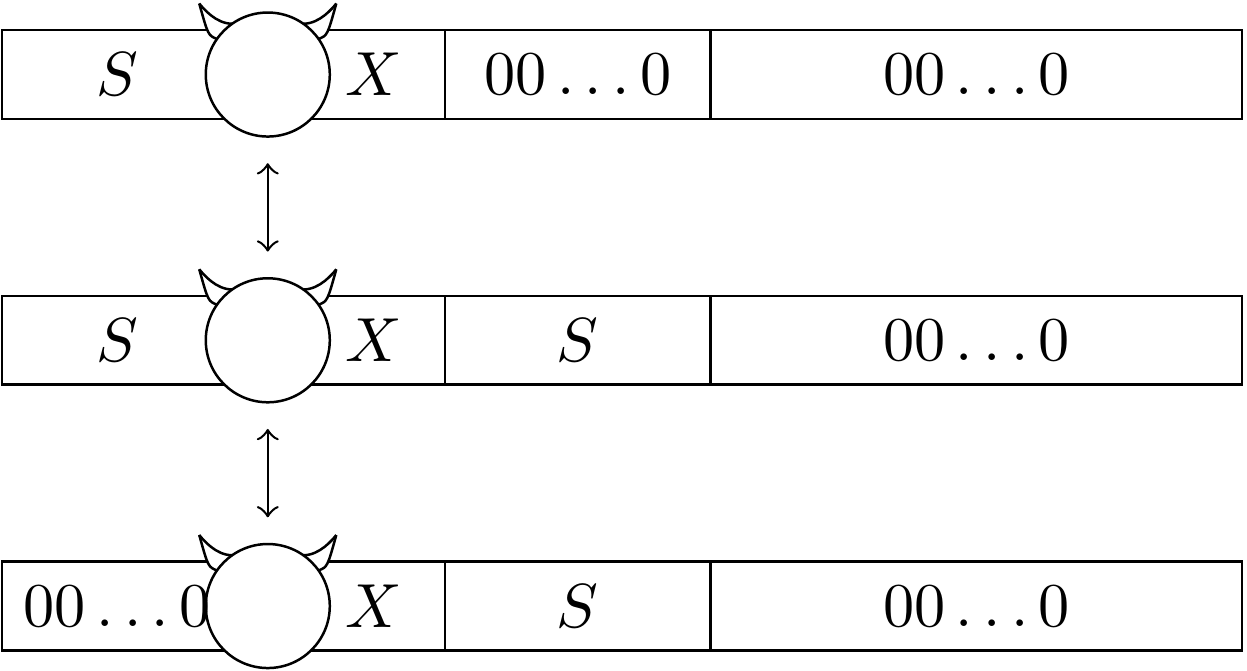}
	\caption{The demon uses~$X$ as program to produce a second copy of~$S$, which thereafter is used to generate~$0^{\text{len}(S)}$ via the reversible operation~$S\oplus S$ (bit-wise addition modulo 2).}
	\label{fig:upperbound}
\end{figure}

Although the same is not true in general, this upper bound is {\em tight\/} if $K_{\cal U}(S|X)\approx 0$.
The latter means that~$X$ itself can be seen as a program for generating an additional copy of~$S$.
The demon  can then bit-wisely XOR this extra~$S$ to the original~$S$ (to be work-extracted) on the tape, hereby producing $0^{{\rm len}(S)}$ {\em reversibly\/} to replace the original~$S$, at the same time saving the new one, as  reversibility demands (see Figure~\ref{fig:upperbound}).
When Bennett's ``uncomputing trick'' is used~|  allowing to make any computation by a~Turing machine logically reversible~\cite{Bennett1973}~|, then a history string~$H$ is written to the tape during the computation of $S$ from~$X$ such that after  XORing, the demon can, in a (reverse) stepwise manner, {\em uncompute\/} the generated copy of $S$ and end up in the tape's original state~| except that the original~$S$ is now replaced by~$0^{{\rm len}(S)}$:
This results in a maximal work value matching the  (in that case trivial) upper bound.\footnote{Let us compare  our bounds with the  entropy-based results
of~\cite{Dahlsten2009,Dahlsten2011}: According to the latter, a demon {\em 
knowing $S$ entirely\/}
 is able to extract maximal work: ${\rm WV}(S)\approx
{\rm len}(S)kT\ln 2$. 
What does it mean to ``know~$S$''?
The knowledge can consist of 
(a)~a {\em copy\/} of~$S$, or of (b)~the
 ability to 
{\em compute\/} such a~copy with a given program $P$, or 
 (c)~it can  determine~$S$ uniquely
{\em without\/} providing the ability to  compute it.
The constructive as opposed to the entropic groups of results are in accordance
in the cases (a) and (b) but  {\em in conflict\/} in case~(c): For instance,
assume the demon's knowledge about $S$ is:
{\em ``$S$ equals the first~$N$ bits~$\Omega_N$ of the binary expansion of~$\Omega$,''\/}
where,~$\Omega$ is the so-called halting probability~\cite{Chaitin1975} 
of a fixed
universal Turing machine ${\cal A}$  ({\em e.g.}, the demon ${\cal U}$ itself).  Although
there 
is a short  {\em  description\/} of $S$ in this case, 
and~$S$ is thus 
uniquely determined in an {\em entropic\/} sense, 
it is still incompressible, even given that knowledge: 
$\mbox{$
K_{{\cal U}}(\Omega_n\, |\, \mbox{``It is bits $1$--$n$  of TM ${\cal A}$'s
halting probability''})\approx n
$}$:
No work is extractable according to our upper bound. Intuitively,  this 
gap 
opens up whenever the {\em ``description complexity''\/}  is smaller than
the {\em Kolmogorov complexity}.
(Note that a self-reference argument, called {\em Berry paradox}, shows that the 
notion of ``description complexity'' is  problematic 
and  can never be defined consistently for
all strings.)}

\section{Revising Landauer's  Principle}
\label{sec:revlandauer}
Here, we revise Landauer's principle to give a lower and an upper bound on the erasure cost.

\subsection{Connection to Work Value}
For a string $S\in\{0,1\}^N$, let WV$(S|X)$ and EC$(S|X)$ be its work
value and erasure costs, respectively, given an additional string $X$
(a ``catalyst'' which remains unchanged, as above). Then
\begin{align}
	{\rm WV}(S|X)+ {\rm EC}(S|X)=NkT\ln 2\ .
	\label{eq:wvec}
\end{align}

To see this, consider first the combination extract-then-erase.
In the extraction process we gain~WV$(S|X)$ of free energy, and consequently have to erase~$N$ bits.
Since this is {\em one specific way\/} of erasing, we have 
\begin{align}
	{\rm EC}(S|X)\leq NkT\ln 2- {\rm WV}(S|X)\ .
\end{align}
If, on the other hand, we consider the combination erase-then-extract, this leads to 
\begin{align}
	{\rm WV}(S|X)\geq NkT\ln 2- {\rm EC}(S|X)
	\,:
\end{align}
We spend~EC$(S|X)$ of free energy to erase the string and use all of the string as ``fuel.''

\subsection{Bounds on the Erasure Cost}
Given the results on the work value above, as well as the connection 
between the work value and erasure cost, we obtain the following
bounds
on the thermodynamic cost of {\em erasing a string}~$S$ by a demon,
modeled as a universal Turing machine ${\cal U}$ with initial 
tape content~$X$.
\\ \

\noindent
{\bf Landauer's principle, revisited.}
{\it 
	Let $\mathcal{Z}$ be a~computable  function,
$
\mathcal{Z}\, :\, \{0,1\}^*\times \{0,1\}^* \longrightarrow \{0,1\}^*
$,
such that 
$
(A,B)\mapsto (\mathcal{Z}(A,B),B)
$
is injective. 
Then we have }
\begin{align}
	K_{{\cal U}}(S|X)kT\ln 2
	&\leq {\rm EC}(S|X)\notag\\
	&\leq {\rm len}(\mathcal{Z}(S,X))kT\ln 2
	\label{eq:revisited}
	\ .
\end{align}
The first inequality follows from Equation~\eqref{eq:wvec} in combination with the upper bound~\eqref{eq:wvupperbound}, the second from Equation~\eqref{eq:wvec} and the lower bound~\eqref{eq:wvlowerbound}.

\section{Generalizing Landauer's  Principle}
\label{sec:gen}
Erasure as well as work extraction can be seen as special cases of a computation with a given input and an output.
Here, we generalize Landauer's principle and discuss the work cost and work value of a general computation, {\it i.e.}, we generalize the already obtained bounds on the cost (minimal amount of free energy that has to be used) and value (maximal amount of free energy that can be gained) to general computation.
Assume that a (universal) Turing machine performs a computation such that the initial content of the tape is $A$ and $X$ (plus a corresponding finite number of $0$s) and the final state is $B$ and $X$ (where $X$ can be seen, again, as a ``catalyst'').
Depending on $A$, $B$, and $X$, this computation can have a work {\em cost\/} or {\em value}, respectively. 
If it has some work cost, then the party performing the computation has to invest free energy that will be dissipated as heat to the environment during the computation.
In the case that the computation has some value, heat from the environment is transformed to free energy.

\subsection{The Energy Cost of a General Computation}
The following result is an algorithmically constructive modi\-fication of 
entropic results~\cite{Faist2015} and a generalization of less
constructive but also complexity-based claims~\cite{Zurek1989a}.

\

\noindent
{\bf  Work cost of a general computation.}
{\it 
	Let $\mathcal{Z}$ be a~computable  function,
$
\mathcal{Z}\, :\, \{0,1\}^*\times \{0,1\}^* \rightarrow \{0,1\}^*
$,
such that 
$
(V,W)\mapsto (\mathcal{Z}(V,W),W)
$
is injective. 
Assume that the Turing machine ${\cal U}$ carries out a computation such
that~$A$ 
is its initial state, $C_1$  the first intermediate state, $C_2$ the
second, etc., up to $C_n$, and $B$ is the final state. 
Then the energy cost  of this computation with side information~$X$
(always on the tape),
${\rm Cost}_{\cal U}(A\rightarrow_{\{C_i\}} B\, |\, X)$, is at least
\begin{align}
	{\rm Cost}_{\cal U} (A\rightarrow_{\{C_i\}}  B\, |\, X)\ \geq\  kT\ln 2\cdot \bigg[K_{\cal{U}}(A|X)\notag\\
	-\sum_{i=1}^n{\big({\rm len}(\mathcal{Z}(C_i,X))-K_{\cal{U}}(C_i|X)\big)}-{\rm len}(\mathcal{Z}(B,X))\bigg]
		\,.
\end{align}
}
\noindent
{\it Proof.}
Let us consider the computation from $(A,X)$ to~$(C_1,X)$.
According to the above (see expression~\eqref{eq:revisited}), the erasure cost of $A$, given $X$, is at least $K_{\cal U}(A|X)\cdot kT\ln2$. 
{\em One\/} possibility of realizing this complete erasure of $A$ is to first transform it to $C_1$ (given $X$), and then erase $C_1$~--- at cost at most len$(\mathcal{Z}(C_1,X))\cdot kT\ln2$.
Therefore, the cost to get from $A$ to $C_1$ given $X$ cannot be lower than the difference between $K_{\cal U}(A|X)\cdot kT\ln2$ and len$(\mathcal{Z}(C_1,X))\cdot kT\ln2$.
The statement follows by summing all contributions of the individual computing steps. 
{\em qed.\/}

\

Note that if no intermediate results are specified, the bound simplifies to
\begin{align}
	{\rm Cost}_{\cal U}&(A\rightarrow B\, |\, X)\ \geq\notag\\
	&[K_{\cal U}(A|X)-{\rm len}(\mathcal{Z}(B,X)) ] \cdot kT\ln2
\end{align}
(see also Ref.~\cite{Wolf2018}).
\begin{figure}[h]
	\centering
	\includegraphics[scale=0.3]{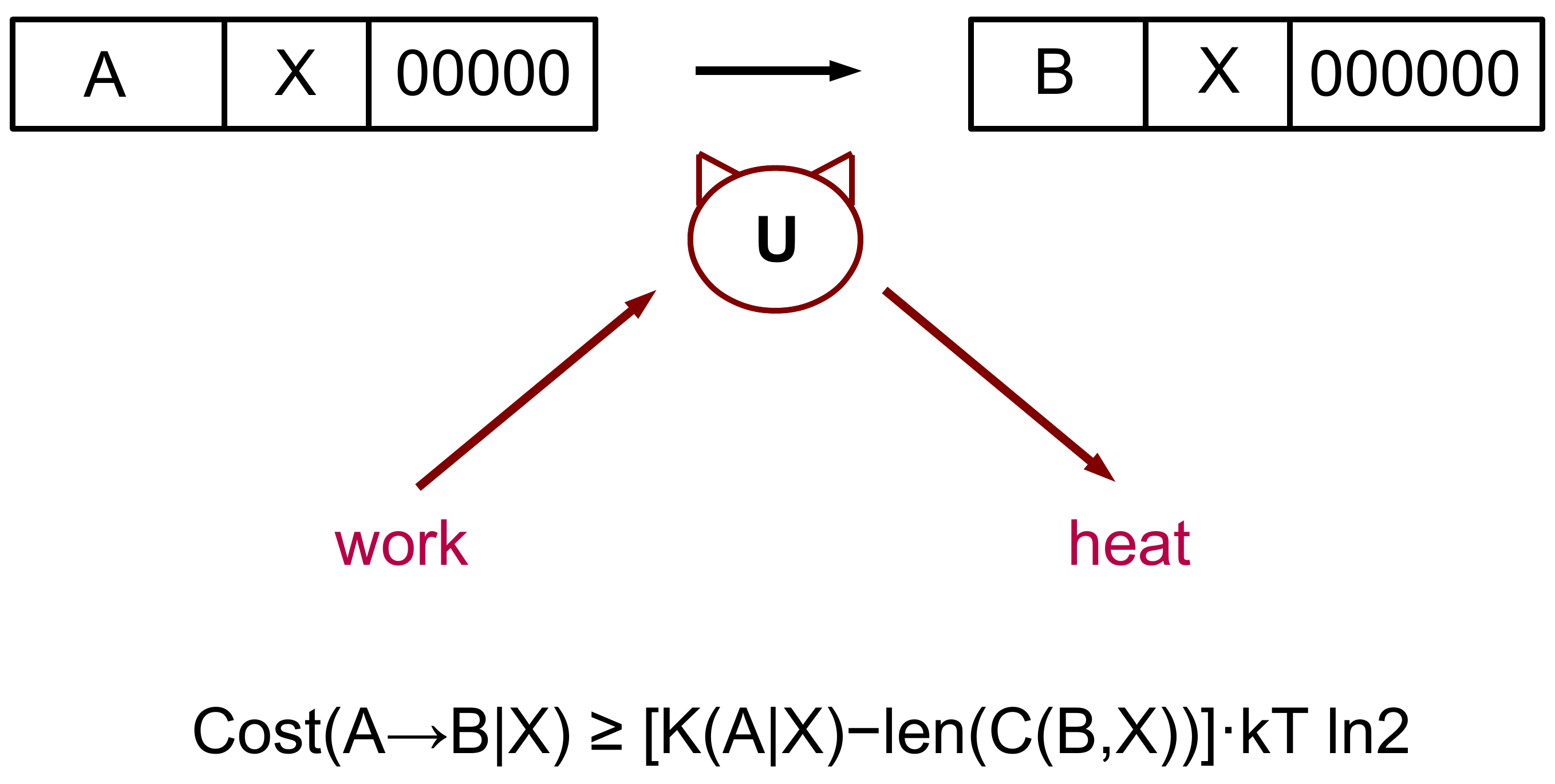}
	\caption{The energy cost of a general computation.}
	\label{fig:3}
\end{figure}

\subsection{The Energy Value of a General Computation}
We consider the {\em work value of a computation\/} from $A$ to~$B$,
given $X$. More specifically, this is a computation that starts with
$(A,X)$ and finishes with $(B,X)$, where $B$ is {\em freely choosable\/} by the
computation among all strings with a given complexity
$K_{\cal{U}}(B|X)$. The work value is denoted  by
${\rm WV}_{\cal U}(A\rightarrow B\, |\, X)$, and it is bounded from
below as follows.

\

\noindent
{\bf Work value of a general computation.}
{\it 
	Let $\mathcal{Z}$ be a computable  function,
$
\mathcal{Z}\, :\, \{0,1\}^*\times \{0,1\}^* \rightarrow \{0,1\}^*
$,
such that 
$
(V,W)\mapsto (\mathcal{Z}(V,W),W)
$
is injective. 
The work value of a~computation from $A$ to $B$, given $X$, is bounded
from below by
\[
	{\rm WV}_{\cal U} (A\rightarrow B\, |\, X)\ \geq\  [K_{\cal U}(B|X)-{\rm len}(\mathcal{Z}(A,X)) ]
\cdot kT\ln2\ .
\]
}\noindent
{\it Proof.}
The cost of erasing $A$, given $X$, is at most len$(\mathcal{Z}(A,X))\cdot kT\ln 2$ (see expression~\eqref{eq:revisited}).
We use a stretch of the resulting all-$0$-string of some length~$N$ for
gaining $NkT\ln 2$ free energy. The resulting string of length $N$ is
then used as a program for the universal Turing machine ${\cal U}$,
with additional input~$X$, and where the computation is made logically
reversible using Bennett's ``uncomputing'' trick~\cite{Bennett1973}; let~$B$ be the resulting string. Then
$K_{\cal{U}}(B|X)\leq N$.
{\em qed.\/}

\subsection{Combination}
Consider the following ``circular computation,'' given~$X$:
\begin{align}
	A\longrightarrow B\longrightarrow A\ .
\end{align}
The free-energy gain of computing $B$ from $A$ is at least 
\begin{align}
	[K_{\cal U}(B|X)-{\rm len}(\mathcal{Z}(A,X))]\cdot kT\ln 2\ ,
	\label{arebeyt}
\end{align}
whereas the cost for computing $A$ back from $B$ is at least this
same amount.
The identity of the two bounds is not very surprising;
it implies that the bound on the work value is ``at least as tight'' as
the one for the cost of the inverse computation, since otherwise a {\em perpetuum mobile of the second kind\/} results.

\section{``Randomness'' and Quantum Correlations}
\label{sec:quantum}
Landauer's revised principle suggests that the erasure cost of a piece of information is an {\em intrinsic, context-free, physical measure for its randomness\/} independent of probabilities and counterfactual statements (that ``some value {\em could\/} just as well have been {\em different\/}'').\footnote{Moreover, such a point of view allows one to discuss {\em randomness\/} on operational grounds.}
This can be tested in a context in which randomness is central:
{\em Bell correlations\/}~\cite{Bell1964} predicted by quantum theory. 
In a proof of principle, it was shown~\cite{Wolf2015} that in essence, a similar mechanism as in the probabilistic setting arises:
{\em If\/}~the correlation is non-local, the inputs are incompressible, and non-signaling holds, {\em then\/} the outputs must be highly complex as well.

Before we describe some of the findings of Ref.~\cite{Wolf2015} in more detail, we introduce the required notation.
For an infinite string~$a=(a_1,a_2,\dots)$, we define its ``truncation''~$a_{[n]}:=(a_1,a_2,\dots,a_n,0,0,\dots)$: the string~$a$ where all symbols after the~$n$-th are set to~$0$.
The expressions~$K(a)$ and~$K(a \mid b)$, where~$b$ is an infinite string as well, denote the functions
\begin{align}
	&K(a):\mathbb{N}\rightarrow\mathbb{N}\,,\text{with}\quad n\mapsto K\left(a_{[n]}\right)\,,\\
	&K(a \mid b):\mathbb{N}\rightarrow\mathbb{N}\,,\text{with}\quad n\mapsto K\left(a_{[n]} \mid b_{[n]}\right)
	\,.
\end{align}
An {\em incompressible\/} string~$a$ has the property
\begin{align}
	K(a) \approx n :\Longleftrightarrow \lim_{n\rightarrow\infty}\frac{K\left( a_{[n]} \right)}{n}=1
	\,,
\end{align}
and a {\em computable\/} string~$a$ the property
\begin{align}
	K(a) \approx 0 :\Longleftrightarrow \lim_{n\rightarrow\infty}\frac{K\left( a_{[n]} \right)}{n}=0
	\,.
\end{align}
Intuitively, the shortest program that prints an incompressible string consists of that very same string, and the shortest program that prints a computable string has a constant length.
Moreover, we say two functions~$f$ and~$g$ mapping natural numbers to natural numbers, where~$g\not\approx 0$, satisfy~$f\approx g$ if and only if
\begin{align}
	\lim_{n\rightarrow \infty} \frac{f(n)}{g(n)} = 1
	\,.
\end{align}

Having this notation at hand, the result stated above is the following.
Let~$(a,b,x,y)$ be a tuple of four infinite binary strings, where
\begin{enumerate}
	\item the PR-box condition~\cite{Popescu1994} is satisfied, {\it i.e.},
		\begin{align}
			x_i\oplus y_i = a_ib_i \text{ for all }i\in\mathbb{N}
			\,,
		\end{align}
	\item both ``input'' strings~$a$ and~$b$ are independent and incompressible, {\it i.e.},~$K(a,b) = K(a)+K(b)\approx 2n$,
	\item the no-signaling condition is satisfied, {\it i.e.},
		\begin{align}
			&K(x\mid a)\approx K(x\mid a,b)\,,\text{ and}\\
			&K(y\mid b)\approx K(y\mid a,b)
			\,.
		\end{align}
\end{enumerate}
It follows from these conditions that the ``output'' strings are {\em not computable\/} --- even if conditioned on the ``input:''
\begin{align}
	K(x \mid a) = \Theta(n)\text{ and }K(y \mid b)=\Theta(n)
	\,.
\end{align}
Whilst this stated result assumes the existence of correlations not attainable by quantum means~\cite{TsirelsonsBound}, the same article proves an analogous statement for quantum correlations, {\it e.g.}, for the quantum violations of the chained Bell inequalities~\cite{Barrett2005,Colbeck2012}.

These results allow for a discussion of quantum correlations without the usual counterfactual arguments used in derivations of {\em Bell inequalities\/} (combining in a single formula results of different measurements that cannot actually be carried out together). 
Furthermore, this potentially opens the door to novel functionalities, namely {\em complexity amplification and expansion\/}~\cite{Baumeler2017}. 
What results is an {\em all-or-nothing flavor of the Church/Turing hypothesis\/}~\cite{Wolf2017}: Either no physical computer exists that is able to produce non-Turing-computable data | or even a ``device'' as simple as a single photon can.

\section{The Second Law as Logical Reversibility} 
\label{sec:logrev}
In Landauer's principle, the price for the {\em logical\/} irreversibility of the
erasure transformation comes in the form of 
 a {\em thermodynamic\/} effort. (Since the amount of the required free
  energy, and heat dissipation, is proportional  to the length of the
  best  compression of the string, the latter can be seen as a {\em
    quantification\/}
of the erasure transformation's irreversibility.)
In an attempt to harmonize this somewhat {\em hybrid\/} picture, 
we invoke
Wheeler's~\cite{Wheeler1989}
{\em ``It from Bit\/}: Every  {\em it\/}~| every particle, every field of force, even
  the spacetime continuum itself~| 
derives its function, its meaning, its very existence entirely [...]\ from the apparatus-elicited 
answers to yes-or-no questions, binary choices, {\em bits}.''
This is an anti-thesis to Landauer's slogan, and we
propose the following {\em synthesis\/} of the two:  
If Wheeler motivates us to look at the environment as being 
{\em a~computation\/} as well, then Landauer's principle may be
 read as: The necessary environmental compensation for 
the logical irreversibility of the erasure of $S$ is such that 
{\em the overall computation, including the environment, is
logically reversible: no information  ever gets completely lost.}
\\

\vspace{1cm}
\begin{center}
\noindent
{\bf Second law, logico-computational version.}\\
{\it 
Time evolutions of closed systems are injective:\\ Nature computes with
Toffoli, but no AND or OR gates.
}
\end{center}
\ \\
Note that this fact is {\em a priori a\/}symmetric in time:
The future must uniquely determine the past, not necessarily {\em vice versa}.
In case the condition holds also for the reverse time direction, the computation is called {\em deterministic}, and {\em randomized\/} otherwise.

Logical reversibility is a simple computational version of~a
discretized second law,
and it has implications resembling the traditional
versions of the  law:
First of all, it leads to~a ``Boltzmann-like'' form, {\em i.e.}, the
existence of a  quantity essentially monotonic in time. More
specifically, 
the logical reversibility of 
time evolution implies that the Kolmogorov complexity of the global state
at time $t$ can be smaller than the one at time $0$ 
only by at most $K(R_t)+O(1)$ if $R_t$ is a string encoding 
the time span $t$. The reason is that one possibility of describing 
the state at time~$0$ is to give the state at time~$t$, plus~$t$
itself; the rest is exhaustive search using only a~constant-length
program
simulating forward time evolution (including possible randomness). 

Similarly, logical reversibility also implies
statements resembling the version of the second law due to {\em
  Clausius\/}: ``Heat does not spontaneously flow from cold to hot.''
The rationale here
is explained with a toy example: If we have a circuit  | the time
evolution | using only (logically
reversible) Toffoli gates, then it is
{\em impossible\/} that this circuit computes  a transformation 
mapping a~pair of strings to another pair  such that
the 
Hamming-heavier of the two becomes even heavier while the lighter gets lighter.\footnote{The Hamming weight of a binary string~$S$ is the number of~$1$s in~$S$.}
A~function {\em accentuating  imbalance}, instead of lessening it, is not injective, as the following counting argument shows. 
\\ \

\noindent
{\it ``Clausius'' Toy Example.}
Let a circuit consisting of only Toffoli gates map an $N(=2n)$-bit string to another.
We consider the map separately on the first and second halves and  assume the computed function to be conservative, {\em i.e.},~to leave the Hamming weight of the full string unchanged at~$n$ (conservativity can be seen as some kind of {\em first\/} law, {\em i.e.}, the preservation of a quantity).
We  look at the excess of~$1$s in one of the halves (which equals the deficit of $1$s in the other).
We observe that the probability (with respect to the uniform distribution over all strings of some Hamming-weight couple $(wn,(1-w)n)$, where the first half has~$wn$ 1s and the second~$(1-w)n$) of the {\em imbalance substantially growing\/} is exponentially weak.
The key ingredient for the argument is the function's  injectivity.
Explicitly, the probability that the weight couple goes from~$(wn,(1-w)n)$ to~$((w+\Delta)n,(1-w-\Delta)n)$ | or more extremely |, for $1/2\leq w<1$ and $0<\Delta\leq 1-w$, is
\begin{align}
	\frac{{n \choose (w+\Delta)n}{n \choose (1-w-\Delta)n}}
	{{n \choose wn}{n \choose (1-w)n}}
	=2^{-\Theta(n)}\ .
\end{align}

\

The example suggests that logical reversibility
might be the ``Church/Turing manifestation'' of the second law: If~reality is 
computed by a Turing machine, then physical laws correspond to 
properties of such computations~--- as in the case of the second law: logical
reversibility.
If we assume for a moment that the second law of thermodynamics has indeed  such a simple 
Church/Turing manifestation, it is a natural question inhowfar just
this already
makes the law special. In fact, the law {\em
  does\/} have a  peculiar related property, {\em
  encoding independence\/}: Since the second law deals with degrees of
freedom, and a degree of freedom will correspond, in another encoding, 
to a degree of freedom again, it is either respected 
in both encodings or violated in both. 
(In comparison: It cannot be decided by just looking at a running
program
whether the simulated system ``violates or respects Kepler's
laws''~--- that would
crucially depend on how masses and their position are represented by the code.)
Hand in hand with this comes
the property of {\em simulation resilience}. Let us take again the
example of Kepler's laws: If a program simulates planets moving
on, say, {\em square\/} orbits, this does not mean that the program execution,
viewed as a physical process, 
is by itself  problematic~--- the laws of gravity
are {\em not\/} simulation resilient. If, in sharp contrast to this,
we simulate a system, {\it e.g.},  the microstate sequence of a~steam
engine, then that
simulation process
violates the second law just as the simulated system does: {\em With respect to the second law of
  thermodynamics, 
the simulation of ``reality'' is just as good as  ``reality'' itself}. 

\section{Conclusion}
We start from Landauer's principle, stating that the erasure of
information requires an amount of free energy, to be  dissipated as
heat into the environment,
proportional to the number of independent binary degrees of freedom of that 
information. Specifically, the use of reversible data compression, imagined to be
carried out by 
 Fredkin and Toffoli's ballistic computer, implies that the necessary
 amount 
is proportional to the length of the best compression of the
information into a binary string (and not to the length of the original 
string, as  often stated). We  generalize and broaden the scope
of  the principle, and 
its converse, to lower bounds on the free-energy cost of~--- or gain from~---
{\em a general computation\/}: the  bounds 
on cost {\em versus\/} gain are in accord.

Landauer has derived, in 1961, his principle from the second law
of thermodynamics. We close the circle by formulating a simple 
``Church/Turing version'' of that law: {\em Logical reversibility
of the overall computation, including the environment}. This fact 
alone implies  variants of the historical versions of 
the second law, due to Boltzmann, Clausius, also Kelvin; it is perhaps 
equivalent to them, certainly  
   simpler. 
The  arising belief that  the law is rather {\em
  logical\/} than {\em physical\/} in its nature is nourished by 
two properties of the second law: {\em its
  encoding independence\/} and its {\em 
simulation resilience}.

Confronted with the relevance of the second law of thermodynamics in
computation, and with its {\em simulation resilience}, 
 let us close with the (provocative)
question whether Landauer's {\em ``Information is physical''\/} should
be 
 replaced by
\vspace*{0.2cm}
\begin{center}
		{\bf ``The second law of thermodynamics is not physical.''}
\end{center}
\vspace*{1cm}

\noindent
{\bf Acknowledgments.}
We thank
Charles B\'edard,
Claus Beisbart,
Sophie Berthelette,
Paul Boes,
Gilles Brassard,
\v{C}aslav Brukner,
Xavier Coiteux-Roy,
Bora Dakic,
Paul Erker,
J\"urg Fr\"ohlich,
and Arne Hansen
for enlightening discussions, and anonymous reviewers for helpful comments.
\"A.B.~is supported by the Swiss National Science Foundation (SNSF) under grant 175860, the Erwin Schr\"odinger Center for Quantum Science \& Technology (ESQ), and the Austrian Science Found (FWF): Z3.
S.W.~is supported by the Swiss National Science Foundation (SNSF), the NCCR {\em QSIT}, and the {\em Hasler Foundation}.

\end{document}